\ifx\LaTeXe\undefined
  \documentstyle[12pt,epsfig]{article}
  \newcommand{\ifig}[1]{\mbox{\epsfig{file=#1,height=8cm,width=12cm}}}
\else
  \documentclass[12pt]{article}
  \usepackage{a4}
  \usepackage[dvips]{graphicx}
  \usepackage{latexsym}
  \usepackage{euscript}
  \newcommand{\ifig}[1]{\includegraphics[height=8cm,width=12cm]{#1}}
\fi
\def\bc{\begin{center}}
\def\ec{\end{center}}
\def\be{\begin{equation}}
\def\ee{\end{equation}}
\def\bea{\begin{eqnarray}}
\def\eea{\end{eqnarray}}

\def\simge{\ \lower-
1.2pt\vbox{\hbox{\rlap{$>$}\lower5pt
\vbox{\hbox{$\sim$}}}}\ }

\newcommand{\AC} {{\cal{A}}}
\begin{document}
\pagestyle{empty} 
\vspace{-0.6in}
\begin{flushright}
ROME prep. 1200/98 \\
SNS/PH/1998-004\\
%CERN-TH/98-??? \\
%hep-ph/98???? \\
%TUM-HEP-279/97 
\end{flushright}
\vskip 2.0in

\centerline{\large {\bf{On the Definition of Gauge Field Operators}}}
\centerline{\large {\bf {in}}}
\centerline{\large {\bf {Lattice Gauge-Fixed Theories}}}
\vskip 1.0cm
\centerline{L. Giusti$^{(1,2)}$, M. L. Paciello$^{(3)}$, S. Petrarca$^{(3,4)}$,
B. Taglienti$^{(3)}$, M. Testa$^{(3,4,5)}$}

\centerline{\small $^1$  Scuola Normale Superiore, 
P.zza dei Cavalieri 7, I-56100 Pisa Italy.}
\centerline{\small $^2$INFN, Sezione di Pisa, San Piero a Grado, I-56100
Pisa Italy.}
\centerline{\small $^3$INFN, Sezione di Roma 1,
P.le A. Moro 2, I-00185 Roma, Italy.}
\centerline{\small  $^4$ Dipartimento di Fisica, Universit\`a di Roma "La
Sapienza",}
\centerline{\small P.le A. Moro 2, I-00185 Roma, Italy.}
\centerline{\small $^5$  Theory Division, 
CERN, 1211 Geneva 23, Switzerland$^{\star}$.}

\centerline{\small }
\vskip 1.0in
\abstract{We address the problem of defining  the four-potential, $A^a_\mu(x)$,
on the lattice,
in terms of the natural link variables, $U_\mu(x)$.

Different regularized definitions are shown, through non perturbative
numerical computation, to converge towards the same continuum renormalized
limit.}
\vskip 1.0in
\begin{flushleft} 
%CERN-TH/98-??? \\
ROME prep. 1200/98 \\
SNS/PH/1998-004\\
%TUM-HEP-297/97 \\
%March 1998
\end{flushleft}
\vfill
\noindent \underline{\hspace{2in}}\\
$^{\star}$ Address until August 31st, 1998.

\eject
\pagestyle{empty}\clearpage
\setcounter{page}{1}
\pagestyle{plain}
\newpage 
\pagestyle{plain} \setcounter{page}{1}

\newpage

\section{Introduction}

Lattice QCD does not require, in itself, any gauge-fixing in order to
compute physical quantities. However, as became increasingly clear in
recent years, lattice gauge-fixing provides a necessary instrument in the
study of quantities, like the quark and gluon propagators,
whose behavior could be relevant to the study of the confinement
mechanism\cite{mand_g}-\cite{bern_q}.

Gauge-fixing is also a necessary ingredient in some non-perturbative
renormalization schemes\cite{NPM,parrinello} and it has been shown to
facilitate the construction of composite fermion operators
with correct chiral behavior\cite{NPM}, essential for the study of
hadronic weak interaction phenomenology\cite{eps'/eps}-\cite{BK}.

Once gauge-fixing has been performed, Green's functions containing the
fundamental quark and gluon field insertions, become accessible to
non-perturbative study. However, contrary to what happens for quarks,
a natural definition for the gluon field is missing.
In the general field theoretical framework, as checked in perturbation theory,
this is known not to be a real problem: any pair of operators differing from
each other by irrelevant terms, i.e. formally equal up to terms of order
$a$, will tend, to the same continuum operator, up to a constant.
It is the purpose of this paper to show that this feature is also
satisfied at the non-perturbative level, within the framework
of lattice QCD. We will, in fact, show that different definitions
of the gluon field, at the lattice level, give rise to Green's functions
proportional to each other, thus guaranteeing the uniqueness of the renormalized
continuum gluon field.

The plan of the paper is as follows. In section \ref{uno}, we recall some
basic facts about lattice gauge-fixing. In section \ref{due} we discuss the
ambiguities intrinsic to the lattice definition of gauge potential and
in section \ref{tre} we summarize our results.

\section{Lattice Gauge-Fixing} \label{uno}

The Landau or Coulomb lattice gauge-fixing procedures
are well known and  several numerical algorithms  are available to this
aim.
In the standard approach\cite{mand_g,Davies} the functional:
\be
F [U^{\Omega}] \equiv - \frac{1}{V\cdot T} 
Re \ Tr  \sum_{\mu} \sum_{x} U_{\mu}^{\Omega}(x)
\label{eq:bigf}
\ee
is minimized with respect to $\Omega(x)$. In eq.(\ref{eq:bigf})
$V$ is the lattice spatial volume, $T$ its time extension and
$U_{\mu}^{\Omega}(x) \equiv \Omega(x)
U_{\mu}(x) \Omega(x + \mu)^{\dagger}$ is the compact
$SU(3)$ gauge field, gauge transformed by the local gauge transformation
$\Omega(x)$. The extrema of $F$ with respect to $\Omega$, correspond
to configurations  satisfying the gauge condition
$\partial_\mu A^{\Omega}_\mu = 0$ in discretized form, with the lattice
$A_\mu$ definition:
\be
A_{\mu} (x) \ 
\equiv \ {{( U_{\mu} (x) - U_{\mu}^{\dagger} (x) )_{traceless}}\over
{2 i a g_0}}, \ \  \mu = 1, \ldots 4,
\label{eq:prima}
\ee
where $a$ is the lattice spacing and $g_0$ is the lattice bare
coupling constant.
The numerical behaviour of the gauge-fixing algorithm is usually monitored,
as a function of the number of sweeps, by two quantities.
The first is $F[U^\Omega]$ itself, and decreases monotonically.
The second, denoted by $\theta$, is defined as:
\be
\theta \equiv \frac{1}{V \cdot T} \
\sum_{ x} \theta( x) \equiv \frac{1}{V \cdot T}
\ \sum_{x} Tr \ [ \Delta (x) \Delta^{\dagger} (x)]\; ,
\label{eq:theta}
\ee
where:
\be
\Delta( x) \equiv \sum_{\mu} \ 
( A^\Omega_{\mu} ({ x}) - A^{\Omega}_{\mu} (x - \hat{\mu} ))\; .
\label{eq:landau}
\ee
The functional $F[U^\Omega]$ is a lattice discretization of
$\int d^{4} x \ Tr \ (A_{\mu}^{2})$, while $\theta$ corresponds to the
continuum quantity $\int d^{4} x \ Tr \ ( \partial_{\mu} A_{\mu} )^{2}$. 
By its very definition, $\theta$, as a functional of $\Omega$,
decreases (not strictly monotonically) during the gauge-fixing process,
becoming zero when $F[U^\Omega]$ gets constant: its value controls
the fulfillment of the gauge condition.

The main unsolved problems concerning gauge-fixing stem from
the existence of both continuum and lattice Gribov
copies\cite{Gribov}-\cite{usgrib1}
and the numerical noise that they can generate.

Being a result of discretization effects, the lattice Gribov copies are of a
quite different nature from those of the continuum, related to topological
obstructions of the gauge-fixing condition
$\partial_{\mu} A_{\mu} = 0$ \cite{vanbaal}.

This confusing situation of the numerical gauge-fixing has been already
stressed in the literature \cite{shamiretal}. The real concern is,
of course, the influence that these phenomena may have on the value
of continuum observables, when computed through the intermediary of non gauge
invariant quantities in the schemes referred to above.

\section{The Lattice Gauge Potential} \label{due}

In this section we will discuss the problems related to the
ambiguities in the lattice definition of the gauge potential.

A natural definition of the 4-potential in terms of the
links, $U_{\mu}$, which represent the fundamental dynamical gluon variables,
is given in eq.(\ref{eq:prima}). 
This definition is naively suggested by the interpretation of $U_{\mu}(x)$
as the lattice parallel transport operator and by
its formal expression in terms of
the "continuum" gauge field variables, $ A_{\mu}(x)$ as:
\be
U_{\mu}(x)\equiv \exp(i g_0 a A_{\mu} (x)).
\ee
A formal expansion in powers of $a$, shows that eq.(\ref{eq:prima}) defines
$ A_{\mu}(x)$, up to terms formally of order $a^2$.

It is clear that the definition given in eq.(\ref{eq:prima}) is far from unique:
it cannot be preferred to any other definition with analogue properties
as, for instance:
\be
{A^{'}}_{\mu} (x) \
\equiv \ {{( (U_{\mu} (x))^2 - (U_{\mu}^{\dagger} (x))^2 )_{traceless}}
\over {4 i a g_0}}, \ \  \mu = 1, \ldots 4. \label{eq:seconda}
\ee
which in fact differs from the one given in eq.(\ref{eq:prima})
by terms of $O(a)$ that formally go to zero as $a \rightarrow 0$.

>From the algorithmical point of view, however, the various definitions are
not interchangeable. In fact let us see what happens if we fix the gauge
of a thermalized configuration stopping the gauge-fixing sweeps
when $\theta \leq 10^{-14}$ and then define $\theta^{'}$ as having the
same functional form of $\theta$, as in eq. (\ref{eq:theta}),
with $A_\mu$ replaced by $A^{'}_\mu$.

\begin{figure}
\bc
\ifig{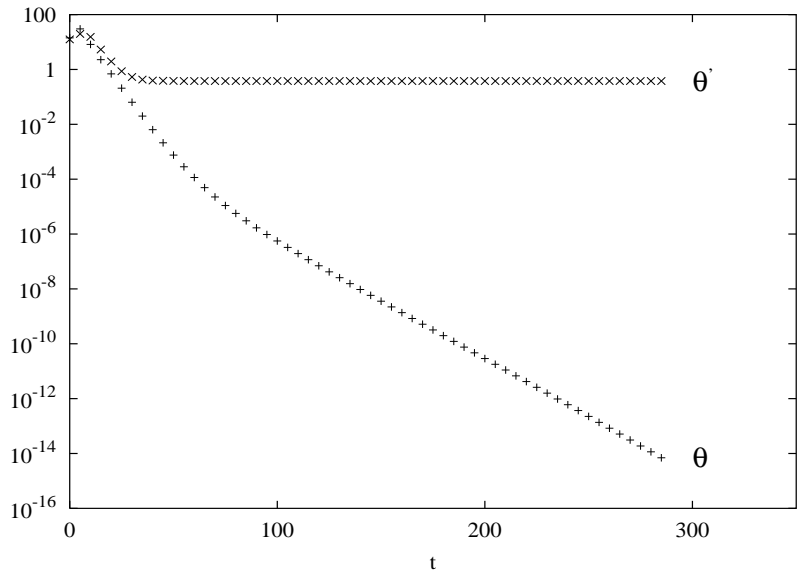}
\caption{\small{Typical behaviour of $\theta$ and $\theta^{'}$ vs gauge fixing sweeps at $\beta=6.0$
for a thermalized $SU(3)$ configuration $8^3\cdot 16$.}}
\label{fig:p60t}
\ec
\end{figure}

The values of $\theta$ and $\theta^{'}$ during the minimization
of $F$ are reported in Fig.\ref{fig:p60t}, for a typical thermalized
configuration, as functions of the lattice
sweeps of the numerical gauge-fixing algorithm. As clearly seen
$\theta^{'}$ does not follow the same decreasing behaviour as $\theta$:
after an initial decrease, $\theta^{'}$ goes to a constant value, many orders of
magnitude higher than the corresponding value of $\theta$.
This fact has been already remarked in ref.\cite{giusti}
where it has been attributed to the large contribution of order $a$ terms.

This marked difference between the behavior of $\theta$ and
$\theta^{'}$ seems to cast doubts on the lattice gauge-fixing
procedure and on the corresponding continuum limit of gauge dependent operators.
On the contrary, we will show, at the end of this section, that
this discrepancy has a natural field theoretical explanation. 

The relation between the two lattice definitions $A_{\mu}(x)$
and $A^{'}_{\mu}(x)$ is of the form:
\be
{A^{'}}_{\mu}(x)= A_{\mu}(x)+ a^2 W_\mu(x) \label{eq:relazione1}
\ee
where the $W_\mu(x)$ is a dimension $3$ operator\footnote{We are assuming
that only one operator $W_\mu(x)$ is present, in order to simplify
the presentation. In general several dimension $3$ operators exist and
their mixing has to be taken into account, but the conclusions remain
unchanged.} with the same quantum numbers of
$A_{\mu}(x)$.
The operator $W_\mu(x)$ in eq.(\ref{eq:relazione1}) is not renormalized,
so that, while formally it would seem safe to neglect its contribution
in the continuum limit, in fact we must take it carefully into account.
We start constructing, out of $W_\mu(x)$, a finite, renormalized operator,
$W^R_\mu(x)$ as:
\be
W^R_\mu(x) = Z_W(g_0,a\,\mu_{_R})(W_\mu(x)+ {1-C(g_0) \over a^2}
A_{\mu}(x)) \label{eqw}
\ee
where $\mu_{_R}$ is the mass renormalization scale of the theory and
$Z_{W}$ is a logarithmically divergent, subtraction dependent
constant, while $C$, as a consequence of the Callan-Symanzik equation,
can only depend on the bare coupling $g_0$\cite{prep}.
Eqs.(\ref{eq:relazione1}) and (\ref{eqw}) show that
\be
{A^{'}}_{\mu}(x)= C(g_0)A_{\mu}(x)+ {a^2 \over Z_W} W^R_\mu(x)
\ee
so that, up to terms truly of order $a^2$:
\be
{A^{'}}_{\mu}(x)=C(g_0) A_{\mu}(x) \label{eq:relazione}
\ee
For Green's functions insertions we have, therefore, in general:
\be
{{\langle \dots A'_{\mu}(x) \dots \rangle} \over {\langle \dots A_{\mu}(x)
\dots \rangle}}= C(g_0) \label{green}
\ee 
We have numerically checked some consequences of eq.(\ref{green})
by measuring on different
lattices, whose characteristics are reported in Table~\ref{tab:params}, the
following Green functions for $SU(3)$ in the Landau gauge with periodic
boundary conditions:
\begin{eqnarray}
\langle {\AC}_0{\AC}_0\rangle (t) &\equiv & \frac{1}{V^2}
\sum_{{\bf x},{\bf y}} Tr \langle  A_0({\bf x},t)A_0({\bf y},0) \rangle 
\label{eq:A0A0}\\
\langle \AC_i\AC_i\rangle (t) &\equiv & \frac{1}{3 V^2}  
 \sum_{i}\sum_{{\bf x},{\bf y}} Tr \langle  A_i({\bf x},t)A_i({\bf y},0)\rangle 
\label{eq:AiAi}\\
\langle \partial \AC \partial \AC \rangle (t) &\equiv &\frac{1}{16 V^2}  
 \sum_{\mu,\nu}\sum_{{\bf x},{\bf y}} Tr \langle  \partial_{\mu} 
 A_{\mu}({\bf x},t)\partial_{\nu}A_{\nu}({\bf y},0) \rangle 
\label{eq:dAdA}\\
\langle \AC_0\partial \AC\rangle (t) &\equiv &\frac{1}{4 V^2}
 \sum_{\mu}\sum_{{\bf x},{\bf y}}
 Tr \langle  A_0({\bf x},t)\partial_{\mu}A_{\mu}({\bf y},0)\rangle 
\label{eq:A0dA}\\
\langle \AC_i \partial \AC\rangle (t) &\equiv & \frac{1}{12 V^2}
  \sum_{\mu,i}\sum_{{\bf x},{\bf y}}
  Tr \langle  A_i({\bf x},t)\partial_{\mu}A_{\mu}({\bf y},0)\rangle 
\label{eq:AidA}
\end{eqnarray}
using both $A$ and $A^{'}$, as defined in eqs.(\ref{eq:prima})
and (\ref{eq:seconda}).
In eqs.(\ref{eq:A0A0})-(\ref{eq:AidA}) the trace is over the color indexes,
$\mu$ and $\nu$ run from 1 to 4 and  $i$ from 1 to 3.
Here and in the following we define $\AC_{\mu}(t)=\sum_{\bf x} A_{\mu}({\bf x},t)$.

The correlators defined in eqs.(\ref{eq:A0A0})-(\ref{eq:AidA}), and
in particular $\langle \AC_i\AC_i\rangle (t)$, are relevant to the
investigation of the QCD gluon sector\cite{mand_g}-\cite{bern_q}.
In this paper we will not be concerned with the interpretation
of their $t$ behavior: our aim is to show their independence on 
the ambiguities related to the definition of the gluon field.

The proportionality factor, $C(g_0)$, may depend on the direction
$\mu$, if the lattice
breaks cubic symmetry. In our case, as shown in Table~\ref{tab:params},
two of the lattices have a time extension different from the spatial one,
so that we have a coefficient $C_0(g_0)$ relating $A'_0$ to $A_0$ and a
different one, $C_i(g_0)$, connecting $A'_i$ to $A_i$.

\begin{figure}
\bc
\ifig{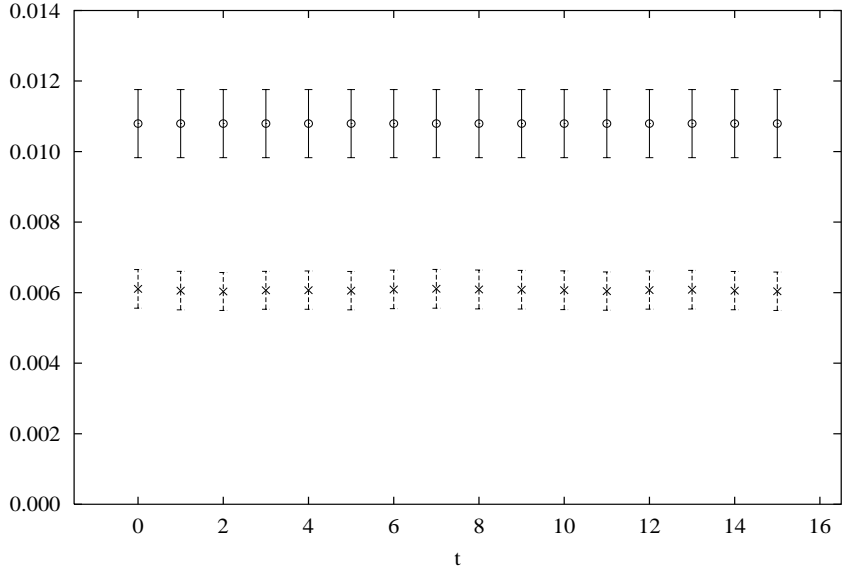}
\caption{\small{Comparison of the matrix elements of $\langle\AC_0\AC_0\rangle(t)$ (open circles) 
and $\langle\AC^{'}_0\AC^{'}_0\rangle(t)$ (crosses) as function of time for a set of 
50 thermalized $SU(3)$  configurations at $\beta=6.0$ with a volume 
$V\cdot T=8^3\cdot 16$ (run $W60b$); the errors are jacknife.}}
\label{fig:a0a0}
\ec
\end{figure}

It is worth noting that $\langle {\AC}_0{\AC}_0\rangle (t)$, when
evaluated through $A_\mu(x)$, should be
constant in $t$ configuration by configuration, in virtue of 
the Landau gauge condition which, together with periodic boundary
conditions, implies $\partial_0 \AC_0 = 0$.
The same should be true, on average, when $A^{'}$ is used.
The behaviour of these two correlators is shown in
Fig. \ref{fig:a0a0} for the run W60b where the errors have been
evaluated through jacknife. Fig.~\ref{fig:a0a0} puts in evidence the
flatness of $\langle \AC^{'}_0\AC^{'}_0\rangle$ and $\langle \AC_0\AC_0\rangle$. 
As remarked above, $\langle \AC_0\AC_0\rangle$ has to be constant configuration
by configuration,
which is verified with a precision of $\sim 10^{-6}$.
More surprising is the fact that also $\langle \AC^{'}_0\AC^{'}_0\rangle$
turns out to be constant configuration by configuration at the level of
$\sim5\%$, also because in this case the behavior of the control variable
$\theta^{'}$, displayed in Fig.~\ref{fig:p60t}, shows that $A^{'}$ is far from
satisfying the Lorentz condition on individual configurations.

For both correlators, the error is just due to fluctuation of 
their constant (in $t$) value, configuration by configuration.

\begin{figure}
\bc
\ifig{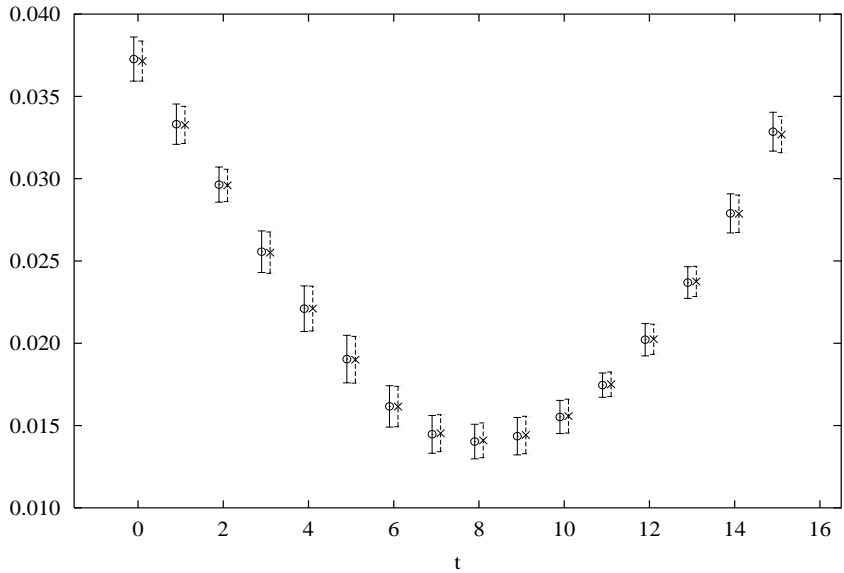}
\caption{\small{Comparison of the matrix elements of  $\langle\AC^{'}_i\AC^{'}_i\rangle(t)$ 
(crosses) and the rescaled $\langle\AC_i\AC_i\rangle~\cdot~C_i^2(g_0)$
(open circles) as function of time for 
a set of 50 thermalized $SU(3)$  configurations at $\beta=6.0$ with a 
volume $V\cdot T=8^3\cdot 16$ (run $W60b$). The data have been
slightly displaced in $t$ to help eye, the errors are jacknife.}}
\label{fig:amu}
\ec
\end{figure}

In Table~\ref{tab:params} we report the fit of the ratio:
\be
{{\langle \AC^{'}_i\AC^{'}_i\rangle} \over {\langle \AC_i\AC_i\rangle}}
\equiv C_i^2(g_0)
\ee
as a constant in time.
In Fig.~\ref{fig:amu} the Green function $\langle \AC^{'}_i\AC^{'}_i\rangle$
and the rescaled one
$C_i^2(g_0) \langle \AC_i\AC_i\rangle$ are reported for the run W60b.
The remarkable agreement between these two quantities
confirms the proportionality shown in eq. (\ref{eq:relazione}).

As shown in Table~\ref{tab:params}, $C_0(g_0)$ and $C_i(g_0)$
coincide, within the errors, for the symmetric lattices W58 and W60a ,
while they have a different value for W60b and W64. This is
probably due to the breaking of cubic symmetry.
This interpretation is confirmed by considering the
W60a ($8^3 \cdot 8$) and W60b ($8^3 \cdot 16$) lattices, with the
same $\beta$ value and configured so that the time extension of W60a
is equal to the spatial extension of W60b. In fact we find that the coefficients
$C_0(g_0)$ evaluated from W60a and $C_i(g_0)$ estimated from W60b
agree within the errors.

The breaking of cubic symmetry, which depends on the ratio $T/(V)^{1/3}$,
could be a potential
source of systematic error in the non-perturbative evaluation of
renormalization constants on asymmetric lattices.

The remaining Green functions,
eqs.(\ref{eq:dAdA}-\ref{eq:AidA}), which
contain the insertion of $\partial_{\mu} A_{\mu}$ 
exhibit the expected behavior: the ones formed with $A_{\mu}$
fluctuate around zero at a level of $10^{-5}$, while those built
through $A'_{\mu}$ fluctuate around zero at a level of
$10^{-2}-10^{-3}$.

We are now ready to show why the discrepancy between
the values of $\theta$, relevant to control the gauge-fixing algorithm,
and the expectation values of $\theta^{'}$, is natural.
In fact we have:
\bea
\theta = a^4 \int d^4x \ (\partial_\mu A_\mu(x))^2\\
\theta' = a^4 \int d^4x \ (\partial_\mu A'_\mu(x))^2 \nonumber
\eea
In other words, $\theta$ and $\theta'$ are proportional to two different
discretizations of the same continuum operator
$(\partial_\mu A_\mu)^2$. 
However, while $\theta$ vanishes configuration by configuration,
as a consequence of the gauge fixing procedure, $\theta'$ is proportional
to $(\partial_\mu A'_\mu)^2$, which has the vacuum quantum numbers and
mixes with the identity. The expectation value of
$(\partial_\mu A'_\mu)^2$, therefore, diverges as ${1 \over a^4}$
so that $\theta^{'}$ will stay finite, as $a \rightarrow 0$.

\section{Conclusions} \label{tre}

In this paper we investigated the problem of the definition of lattice
operators converging, as $a \rightarrow 0$, to the fundamental
continuum gauge fields. 
This construction is affected, at the regularized lattice level,
by an enormous redundancy. However we found convincing
non-perturbative evidence, based on numerical simulations, that this
redundancy will be completely compensated, in the continuum, by
the wave function renormalization needed in order to define
finite Green's functions with gauge field insertions.
Although, on general field theoretical grounds, the validity of such
results is not unexpected, we stress the remarkable fact that it holds
true also in this particular situation in which gauge-fixing is naively
performed, disregarding the problems related to the existence
of lattice and continuum Gribov copies.

We have also discussed and solved the problem of the large discrepancy
between the values of the control functionals $\theta$ and $\theta^{'}$
in terms of renormalization of power-divergent composite operators.

A direction dependent effect in the value of the renormalization
constants has been found on non-cubic lattices, which should caution
against too naive application of infinite volume results to finite
volume lattice numerical data. 

We believe that these features have a general validity and will
survive a more thorough treatment of the gauge-fixing problem.

\section*{Acknowledgements}

Massimo Testa thanks the CERN Theory Division for the kind hospitality.

\newpage

%--------------------------------------------------------------
\setlength{\tabcolsep}{.16pc}
\begin{table}
\begin{center} 
\begin{tabular}{||c|cccc||}
\hline\hline       %1234
&W58&W60a&W60b&W64\\
\hline
$\beta$ & $5.8$       &  $6.0$      &     $6.0$    &   $6.4$      \\
\# Confs&  20         &  100        &     50       &   30         \\
Volume  &$6^3\times 6$&$8^3\times 8$&$8^3\times 16$&$8^3\times 16$\\
\hline
$C_i(g_0)$&    0.689(3)     &   0.729(1)  &   0.729(2)   &   0.757(2)   \\
$C_0(g_0)$&    0.690(7)     &   0.729(1)  &   0.750(1)   &   0.784(2)   \\
\hline
$a^{-1}$ & 1.333(6) & 1.94(5) & 1.94(5) & 3.62(4)  \\ 
\hline
\end{tabular}
\end{center}
\caption{\small{Summary of the lattice parameters used and relative
values of $C_0$ and $C_i$ obtained as discussed in section
\protect\ref{due}. The $a^{-1}$ values are taken from \protect\cite{born,schilling} and are computed
through the string tension.}}
\label{tab:params}
\end{table}
%-------------------------------------------------- 

\end{document}